\documentclass[12pt]{iopart}
\usepackage{graphicx}
\usepackage{cite}

\eqnobysec

\begin{document}

\title[]
{Semi-flexible compact polymers  in two dimensional nonhomogeneous confinement}

\author{D Mar\v{c}eti\'{c}$^1$, S Elezovi\'c-Had\v zi\'c$^2$,  N Ad\v zi\' c$^3$ and I \v{Z}ivi\'{c}$^4$ }

\address{$^1$ Faculty of Natural Sciences and Mathematics, University of Banja Luka, M.~Stojanovi\'{c}a 2, Bosnia and Herzegovina}
\address{$^2$ Faculty of Physics,
University of Belgrade, P.O.Box 44, 11001~Belgrade, Serbia}
\address{ $^3$ Faculty of Physics, University of Vienna, Boltzmanngasse 5, A-1090 Vienna, Austria}
\address{$^4$ Faculty of  Science, University of Kragujevac, Radoja Domanovi\'{c}a 12, Kragujevac, Serbia}
\eads{\mailto {dusanka.marcetic-lekic@pmf.unibl.org}, \mailto{suki@ff.bg.ac.rs}, \mailto{ natasa.adzic@univie.ac.at}, \mailto{ivanz@kg.ac.rs}}

\begin{abstract}
We have studied  the compact phase conformations of semi-flexible polymer chains confined in two dimensional nonhomogeneous media, modelled by fractals that belong to the family of modified rectangular (MR) lattices.  Members of the MR family are enumerated by an integer $p$ $(2\leq p<\infty)$ and fractal dimension of each member of the family is equal to 2.  The polymer flexibility is described by the stiffness parameter $s$, while the polymer conformations are modelled by weighted Hamiltonian walks (HWs). Applying an exact method of recurrence equations we have found that partition function $Z_N$ for closed HWs consisting of $N$ steps scales as $\omega^N \mu^{\sqrt N}$, where  constants $\omega$ and $\mu$  depend on both $p$ and $s$. We have calculated numerically  the stiffness dependence of the polymer persistence length, as well as  various thermodynamic quantities (such as free and internal energy, specific heat and entropy) for a large set of members of MR family. Analysis of these quantities has shown that semi-flexible compact polymers on MR lattices can exist only in the liquid-like (disordered) phase, whereas the crystal (ordered) phase has not appeared. Finally,  behavior of the examined system at zero temperature has been discussed.
\end{abstract}

%\pacs{64.60.ae, 64.60.al, 36.20.Ey, 05.50.+q}

\vskip 5mm


\maketitle

\section{Introduction}
\label{prva}

Behaviour of a linear flexible polymer in various types of solvents has been  extensively  studied  in the past and the subject is well understood, at least when  the universal properties of polymer statistics are under consideration \cite{deGennes}.
The canonical model of a linear polymer is the self-avoiding walk (SAW), which is  a random walk that must not contain self--intersections. It this model, steps of the walk are usually identified with monomers, while the surrounding solvent is represented by a lattice  \cite{Vanderzande}. In a good solvent (high temperature regime) polymer chain is in extended state, whereas in a bad solvent (low temperatures) it is in compact phase. Since in the compact phase a polymer fills up the space as densely as possible, it is often modelled by  Hamiltonian walk (HW), which is a SAW that visits every site of the underlying lattice.

 Most of real polymers, especially biologically important ones, are semi-flexible, but contrary to the flexible polymers,  knowledge of their conformational properties is scarce. The measure of  bending rigidity of a semi-flexible  chain is its persistence length $l_p$, which can be understood as an average length of straight segments of the chain. In a good solvent the stiffness of the polymer only enlarges the  persistence length, while in a bad one (when polymer is compact), an increase of the chain stiffness may promote the transition from a disordered phase (when polymer bends are randomly distributed over the polymer, with finite density) to an ordered crystalline phase (when large rod-like parts of the chain lie in parallel order, with zero density of bends). In order to study  the  compact phase of semi-flexible polymers on homogeneous lattices Flory  introduced a model of polymer melting \cite{Flory56}, in which a compact polymer  is modelled  by HW, while the bending rigidity is taken into account by assigning an extra energy to each bend of the chain. Applying the proposed model within the  mean-field theory, it has been found \cite{Flory56} that there are two compact phases: disordered liquid-like  and ordered  crystal-like phase, and a phase transition caused by competition between the chain entropy and the stiffness of the polymer has emerged. At high temperatures, the entropy dominated disordered phase exists, in which the number of bends in the chain is  comparable with the total number of monomers, and the persistence length is finite. At low temperatures bending energy dominates, so that polymer takes ordered crystalline form, in which bends exist only on the opposite edges of the underlying  lattice. In this phase the persistence length becomes comparable to the lattice size. Using  various techniques, in  a series of papers  \cite{Gujrati80,Saleur,Yoon,Doniach,Irback,Marjolein, Corsi, Jacobsen, Krawczyk}, the existence and  nature of phase transition between these two phases of compact polymers  have been investigated, giving  quite different results for the order of phase transition.

 Besides being interesting from the pure physical point of view, semi-flexible compact polymer models are of great importance for better  understanding of  some biological systems and  processes. For example, DNA condensation~\cite{starostin} and protein folding problem~\cite{febs} take place in squeezed cellular environment and demand for compact states of these rigid polymers. For such systems, a model of fractal ('crumpled') globule for DNA packing in  a chromosome have been proposed \cite{Grosberg} and recently confirmed experimentally \cite{Aiden}.

Hamiltonian walk problem, even in its simplest form, with no interactions involved and on regular lattices, is a very difficult one. Exact enumeration of HWs, which is a prerequisite for further analysis of the compact polymer properties, is limited to rather small lattice sizes. For instance, HWs on $L^2$ square lattice have been enumerated up to size $L=17$~\cite{Jacobsen2007}, and on $L^3$ cube up to $L=4$~\cite{Schram}, which is not sufficient to draw solid conclusions about asymptotic behavior for long compact chains (therefore approximate techniques, such as Monte Carlo algorithms~\cite{Kondev,Jacobsen2008} have been used). In addition to the HWs enumeration, solving the semi-flexible HW problem requires their classification according to the number of bends, which makes it even less feasible.
On the other hand, in real situations polymers are usually situated in nonhomogeneous media, so that models of semi-flexible compact polymers should be extended to such environments. In that sense, as a first step towards more realistic situations, fractal lattices may be used as underlying lattices for semi-flexible HWs. Some deterministic fractal lattices have already been useful in exact studies of flexible HWs~\cite{hwfractals}. In these studies, emphasis has been put on establishing the scaling form of the number of very long walks, which is a long-standing issue in various polymer models~\cite{scaling}. Recently, a closely related problem of finding the scaling form of the partition function of semi-flexible HWs on 3- and 4-simplex lattices has been analyzed~\cite{Phaysica2011} in an exact manner. In this paper, we apply a similar approach for enumeration and classification of, in principle, arbitrarily long semi-flexible HWs, in order to find the partition function, as well as various thermodynamic properties of compact polymers adsorbed on two dimensional nonhomogeneous substrates, represented by fractals from the family of modified rectangular (MR) lattices.

 The paper is organized as follows. In section~\ref{druga} we describe the MR lattices for general scaling parameter $p$, introduce the model of semi-flexible HWs and the method of recurrence relations for exact  evaluation of partition function. In the same section we present specific results obtained for $p=2$ MR lattice, and  we analyze  thermodynamic quantities concerning the studied model. In section~\ref{cetvrta} we expose results for lattice with arbitrary $p>2$. The behavior of the studied polymer model at temperature $T=0$ (ground state) is examined in section~\ref{peta}.
Summary of obtained results and  pertinent conclusions are presented in  section~\ref{sesta}.

\section{Semi-flexible closed Hamiltonian walks on the family of modified rectangular lattices}
\label{druga}

\begin{figure}%[b]
\begin{center}
\includegraphics[scale=0.55]{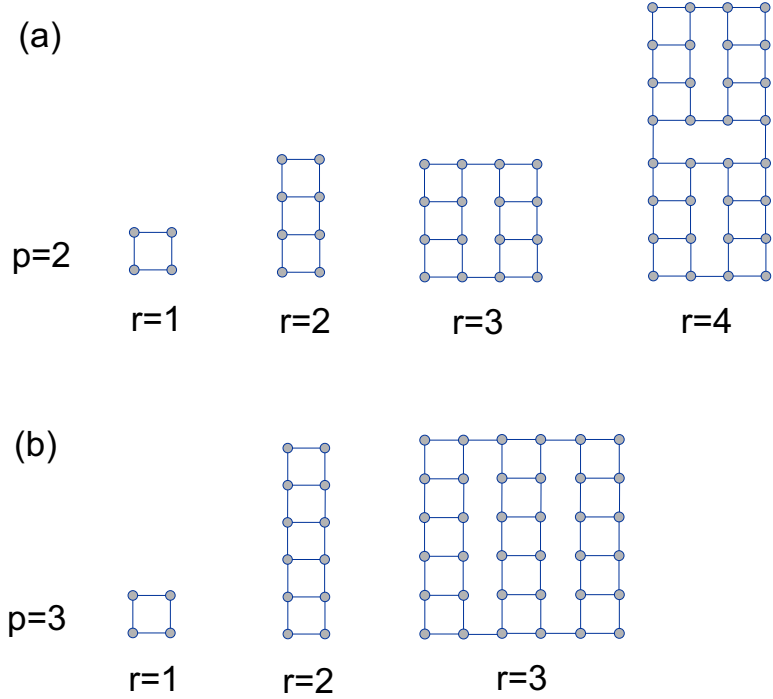}
\end{center}
\caption{(a) First four steps of iterative construction of $p=2$ MR fractal lattice. (b) First three steps in construction of $p=3$ MR fractal.}
 \label{fig:mrlattice}
\end{figure}

In this section  the method of recurrence relations for studying the conformational properties of  compact semi-flexible polymers is described.   Polymer rings  are modeled by  closed HWs (Hamiltonian cycles), whereas the substrates on which the polymers are adsorbed  are  represented by fractals belonging to the  MR family of fractals \cite{Dhar}.  Members of MR fractal family  are labeled by an integer $p$ ($2\le p<\infty$), and can be constructed iteratively.  For each particular $p$, at the first stage ($r=1$) of the construction one has four points forming a unit square. Then, $p$ unit squares are joined in the rectangle to obtain the $(r=2)$ construction stage. In the next step, $p$ rectangles are joined into a square, and so on (see figure~\ref{fig:mrlattice}). The complete lattice is acquired in the limit $r\to\infty$. The lattice structure obtained in the $r$th stage is called the $r$th order fractal generator. It contains $N_{r}=4p^{\,r-1}$ lattice sites,  and  fractal dimension is $d_f=2$ for each fractal of the family.

To take into account the polymer stiffness property, to each bend of the walk we assign  the weight factor $s={\mathrm{e}}^{-\varepsilon/k_BT}$ (stiffness parameter), where $\varepsilon>0$ is the bending  energy, $T$ is the temperature, and $k_B$ is the Boltzmann constant. Varying  $T$ and/or  $\varepsilon$,  the stiffness parameter  can take values in the range  $0\leq s\leq 1$, where two opposite limits $s=0$  and  $s=1$ coincide with a fully rigid  and a fully flexible polymer chain, respectively. To evaluate the partition function one has to sum the weights of all possible polymer conformations $\mathcal{C}_N$ with $N$-steps:  $Z_N=\sum_{\mathcal{C}_N}e^{-E(\mathcal{C}_N)/k_BT}$, where $E(\mathcal{C}_N)=\varepsilon N_b(\mathcal{C}_N)$ is the energy  of an $N$-step conformation having $N_b$ bends.  The above partition function can be written as $Z_N=\sum_{\mathcal{C}_N}s^{N_b(\mathcal{C}_N)}=\sum_{N_b}g_{N,N_b}s^{N_b}$, where $g_{N,N_b}$ is the number of $N$-step conformations with $N_b$ bends  (i.e. degeneracy of  the energy level $\varepsilon N_b$).

\subsection{Recursion relations construction for $p=2$ MR lattice  \label{metodsekcija}}

\begin{figure}%[b]
\begin{center}
\includegraphics[scale=0.8]{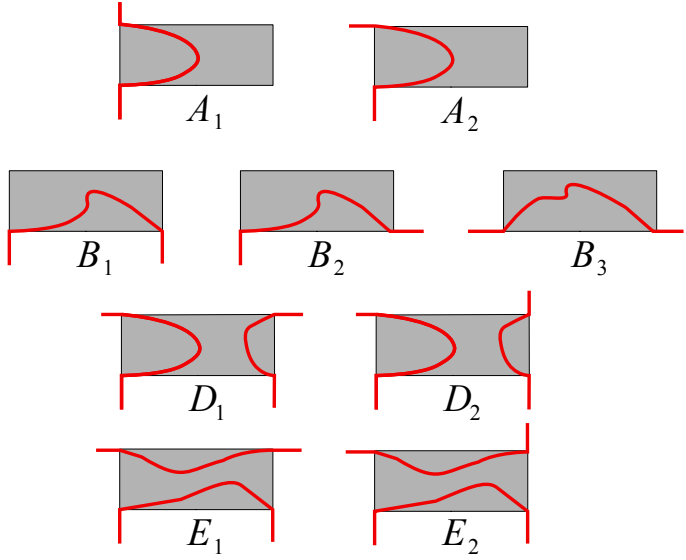}
\end{center}
\caption{Possible types of semi-flexible HWs on the $r$th order lattice structure.}
 \label{fig:redukovane}
\end{figure}

\begin{figure}%[b]
\begin{center}
\includegraphics[scale=0.5]{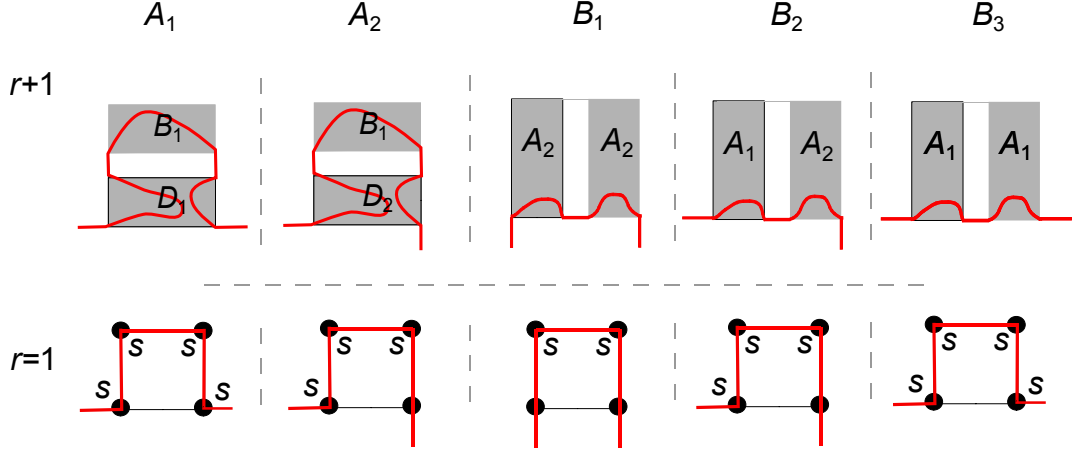}
\end{center}
\caption{Top row: Possible conformations of one-stranded types of semi-flexible HWs on the generator of order $(r+1)$. Gray rectangles represent the $r$th order lattice structure, and curved lines correspond to coarse-grained walks.  Bottom row: Possible one-stranded semi-flexible HWs on the first order generator. Small black circles represent sites the lattice consists of.}
 \label{fig:recur}
\end{figure}

To calculate the partition function for the model under study, one has to enumerate all possible Hamiltonian cycle conformations. In general, this appears to be a very complicated task, but in this case the self-similarity of MR lattices allows systematic enumeration using an exact recursive method~\cite{DusaSM}. In order to explain this approach we present its application in the case of $p=2$ MR lattice.
\begin{figure}%[b]
\begin{center}
\includegraphics[scale=0.5]{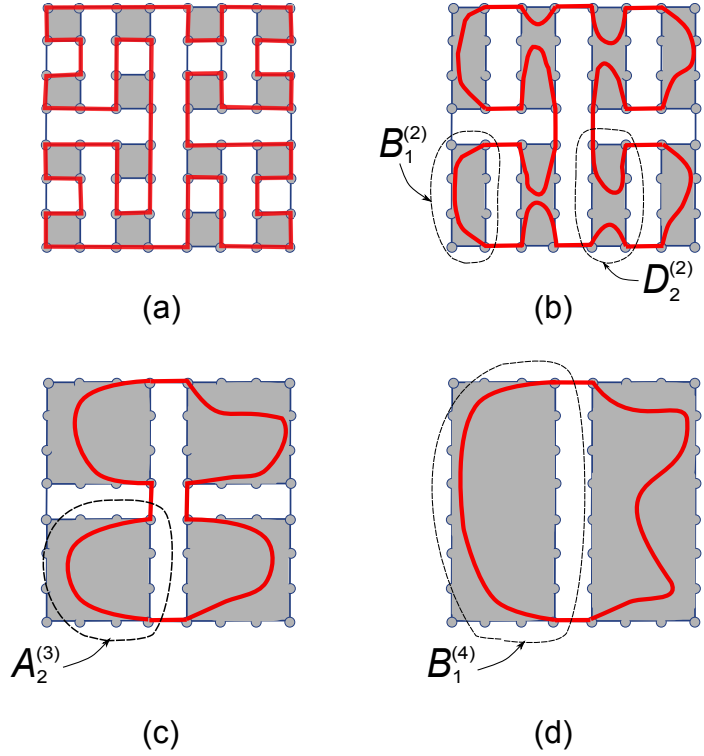}
\end{center}
\caption{(a) Example of a semi-flexible Hamiltonian walk on the 5th order generator of $p=2$ MR lattice. This walk has 42 bends, so that its statistical weight is equal to $e^{-42\epsilon/k_BT}=s^{42}$. Subsequent steps of the coarse-graining process are depicted in (b), (c) and (d). Grey rectangles in (b), (c) and (d) represent generators of order two, three and four, respectively, whereas curved lines correspond to the coarse grained parts of the walk. Different types of conformations within the $r=2$, 3 and 4 generators are encircled. In (d) one can see that this closed Hamiltonian walk, observed on $r=5$ generator, consists of two $B_1$-type Hamiltonian walks which span the two constituent $r=4$ generators. It is obvious that such decomposition of any closed Hamiltonian walk on generator of order $(r+1)$ into the parts within the constituent $r$th order generators is the only possible one.}
 \label{fig:CSFHWMRL}
\end{figure}
In figure~\ref{fig:CSFHWMRL}(a)  an example of closed HW on the $p=2$
MR lattice  of order $r=5$ is shown. Performing a coarse-graining process
 one notices in figure~\ref{fig:CSFHWMRL}(b) that this walk can be
decomposed into several parts corresponding to constitutive second order generators,
which consist of one or two strands. As can be seen in figures ~\ref{fig:CSFHWMRL}(c) and ~\ref{fig:CSFHWMRL}(d), this process can be repeated two more times, leading to a coarse-grained HW consisting of two one-strand parts within the two constituent $r=4$ generators. On the other hand, any one-strand or two-strands HW within any $(r+1)$th order generator can be decomposed into two one-strand or two-strands HW parts within the two constituent $r$th order generators, and due to the self-similarity of MR lattices, such decompositions do not depend on $r$. In order to take into account the semi-flexibility properly, one should observe nine 'traversing' types of conformations: $A_1$, $A_2$, $B_1$, $B_2$, $B_3$, $D_1$, $D_2$, $E_1$ and $E_2$, which are depicted in figure~\ref{fig:redukovane}.
For each of these nine conformations we define the so-called restricted partition function as
\begin{equation}
X^{(r)}(s)=\sum_{N_b} \mathcal{X}_{N_b}^{(r)}s^{N_b}\, ,
\quad X\in \{A_1, A_2, B_1, B_2, B_3, D_1, D_2, E_1, E_2\}\, ,
\label{xvar}
\end{equation}
where $\mathcal{X}_{N_b}^{(r)}$ is the number of HWs of the type $X$ on the $r$th order fractal structure, with $N_b$ bends.
Then, restricted partition functions, for $p=2$ lattice, obey the following recursion relations
\begin{eqnarray}
\fl A_1^{(r+1)}&=&B_1^{(r)}D_1^{(r)}\, , \qquad A_2^{(r+1)}=B_1^{(r)}D_2^{(r)}\, ,   \nonumber\\
\fl B_1^{(r+1)}&=&\left(A_2^{(r)}\right)^2\, , \qquad B_2^{(r+1)}=A_1^{(r)}A_2^{(r)}\, , \qquad B_3^{(r+1)}=\left(A_1^{(r)}\right)^2\, , \nonumber \\
\fl D_1^{(r+1)}&=&2E_2^{(r)}D_2^{(r)}+\left(B_2^{(r)}\right)^2\, ,  \qquad D_2^{(r+1)}=D_2^{(r)}E_1^{(r)}+E_2^{(r)}D_1^{(r)}+B_2^{(r)}B_3^{(r)}\, ,  \nonumber\\
\fl E_1^{(r+1)}&=&\left(D_2^{(r)}\right)^2\, , \qquad E_2^{(r+1)}=D_1^{(r)}D_2^{(r)}\, ,  \label{eq:recur}
\end{eqnarray}
so that, starting with their values for $r=1$: $A_1^{(1)}=s^4$, $A_2^{(1)}=s^3$, $B_1^{(1)}=s^2$, $B_2^{(1)}=s^3$, $B_3^{(1)}=s^4$, $D_1^{(1)}=s^2$, $D_2^{(1)}=s$, $E_1^{(1)}=s^2$, and $E_2^{(1)}=s^3$, for any particular value of $s$ one can, in principle, numerically find the values of the restricted partition functions for very large $r$ values.
In figure~\ref{fig:recur}, construction of recursion relations for $A$- and $B$-type restricted partition functions, together with their initial conditions, is illustrated. In a similar way one can find recursive relations for the two-stranded partition functions, and the corresponding initial conditions.

Due to the fact  that any closed HW on $(r+1)$th order generator of  $p=2$ fractal can only be decomposed into two $B_1$-type HWs within constitutive $r$th order generator (see figure~\ref{fig:CSFHWMRL}(d)), it follows that corresponding partition function $Z_c^{(r+1)}$, for all closed semi-flexible HWs on $(r+1)$th order lattice structure, has the form
\begin{equation}
Z_c^{(r+1)}=\left(B_1^{(r)}\right)^2\, .\label{eq:statsum}
\end{equation}
Iterating restricted partition functions, one can obtain $Z_c$, and consequently explore the thermodynamic behavior of the model. Applying the recursion relations (\ref{eq:recur}) for various values of $s$ (between 0 and 1), one can show that there is a critical value of the bending parameter $s^*=0.7366671$, such that for $s<s^*$ all restricted partition functions tend to 0  (and so does the overall partition function), whereas for $s>s^*$ they all become infinitely large, for $r\gg 1$.  This can be explained by  the  coupling between  the degeneracy $g_{N,N_b}$ of energy levels $E(N_b)=\varepsilon N_b$ and the corresponding  Boltzmann factor $s^{N_b}$.  Degeneracies are  such that they   increase with the energy of levels   attaining their  maximum value,  after which they decrease. At low temperatures  (that is, for small $s$), degeneracies are not large enough to overcome small Boltzmann  factors, but increasing the temperature they prevail and partition function iterates to infinity.

\begin{figure}%[b]
\begin{center}
\includegraphics[scale=0.4]{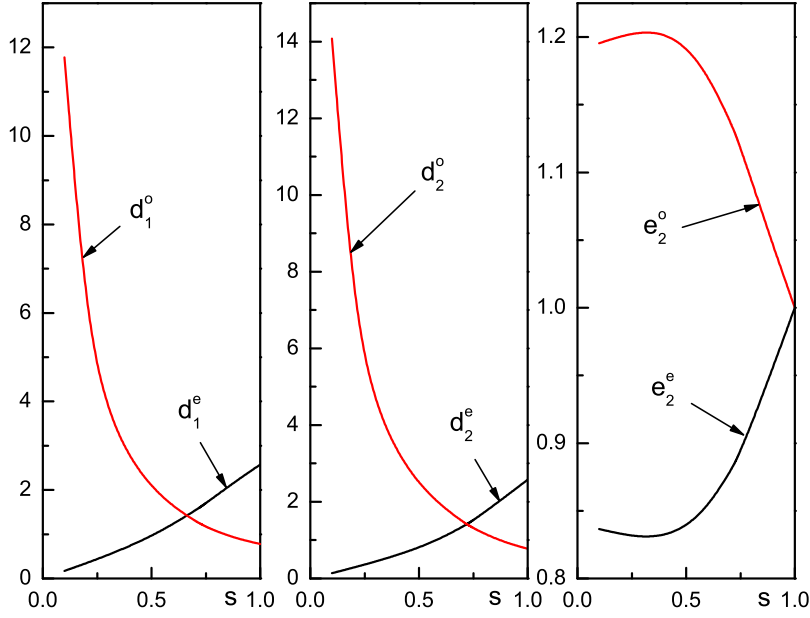}
\end{center}
\caption{Dependance of the limiting values $d_1^{o,e}$, $d_2^{o,e}$ and $e_2^{o,e}$, defined in (\ref{eq:limesi}), on the stiffness parameter $s$, for $p=2$ MR lattice.}
 \label{fig:limesi}
\end{figure}

\begin{figure}%[b]
\begin{center}
\includegraphics[scale=0.38]{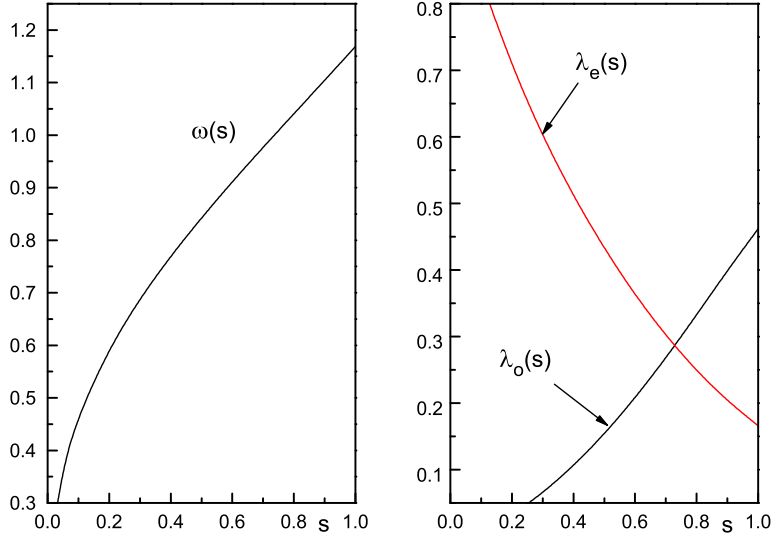}
\end{center}
\caption{Dependence of $\omega$  (\ref{eq:omega}), $\lambda_e$ and $\lambda_o$ (\ref{eq:lambda}) on the stiffness parameter $s$, for $p=2$ MR lattice.}
\label{fig:omegaLambda}
\end{figure}

It order to learn the asymptotic behaviour of partition function (\ref{eq:statsum}), it is   useful
to  introduce  rescaled variables
\begin{equation}
x^{(r)}=\frac{X^{(r)}}{E_1^{(r)}}\, , \qquad x\in \{a_1, a_2, b_1, b_2, b_3, d_1, d_2, e_2\}\, ,  \label{eq:skalirani}
\end{equation}
which fulfil the following recursion relations
\begin{eqnarray}
\fl a_1^{(r+1)}&=&\frac{b_1^{(r)}d_1^{(r)}}{\left(d_2^{(r)}\right)^2}\, , \qquad a_2^{(r+1)}=\frac{b_1^{(r)}}{d_2^{(r)}}\, , \qquad e_2^{(r+1)}=\frac{d_1^{(r)}}{d_2^{(r)}}\, , \nonumber\\
\fl b_1^{(r+1)}&=&\left(\frac{a_2^{(r)}}{d_2^{(r)}}\right)^2\, , \qquad b_2^{(r+1)}=\frac{a_1^{(r)}a_2^{(r)}}{\left(d_2^{(r)}\right)^2}\, , \qquad b_3^{(r+1)}=\left(\frac{a_1^{(r)}}{d_2^{(r)}}\right)^2\, , \nonumber\\
\fl d_1^{(r+1)}&=&2\frac{e_2^{(r)}}{d_2^{(r)}}+\left(\frac{b_2^{(r)}}{d_2^{(r)}}\right)^2\, , \qquad
d_2^{(r+1)}=\frac{1}{d_2^{(r)}}+\frac{e_2^{(r)}d_1^{(r)}+
b_2^{(r)}b_3^{(r)}}{\left(d_2^{(r)}\right)^2}\, ,\label{eq:malerec}
\end{eqnarray}
with the initial conditions
\begin{equation}
\fl a_1^{(1)}=b_3^{(1)}=s^2\, , \qquad a_2^{(1)}=b_2^{(1)}=e_2^{(1)}=s\, , \qquad b_1^{(1)}=d_1^{(1)}=1\, , \qquad d_2^{(1)}=s^{-1}\,  .
\end{equation}
Numerical analysis of  (\ref{eq:malerec}) reveals that, for any $s$ in the region $0<s\leq 1$, variables $a_i^{(r)}$ and $b_i^{(r)}$ quickly tend to 0, whereas $d_1^{(r)}$, $d_2^{(r)}$ and $e_2^{(r)}$ (depending on the parity of $r$), tend to some finite non-zero values. In particular, one obtains
\begin{eqnarray}
\fl \lim_{k\to\infty}d_1^{(2k+1)}(s)&=&d_1^o(s)\, , \qquad  \lim_{k\to\infty}d_2^{(2k+1)}(s)=d_2^o(s)\, ,  \qquad \lim_{k\to\infty}e_2^{(2k+1)}(s)=e_2^o(s)\, ,  \nonumber\\
\fl \lim_{k\to\infty}d_1^{(2k)}(s)&=&d_1^e(s)\, , \qquad  \lim_{k\to\infty}d_2^{(2k)}(s)=d_2^e(s)\, ,  \qquad \lim_{k\to\infty}e_2^{(2k)}(s)=e_2^e(s)\, , \label{eq:limesi}
\end{eqnarray}
where dependance of the limiting values $d_1^{o,e}$, $d_2^{o,e}$ and $e_2^{o,e}$ on $s$ is depicted in figure~\ref{fig:limesi}.
Furthermore, the following relations are satisfied
\begin{equation}
d_2^o d_2^e=d_3^o d_3^e=2\, , \qquad e_2^o e_2^e=1\, ,
\end{equation}
so that using relations~(\ref{eq:malerec}), for large $r$ one obtains asymptotic recursion relation
\begin{equation}
b_1^{(r+2)}\approx \frac 14 \left(b_1^{(r)}\right)^2\, ,\end{equation}
which implies that
\begin{equation}
b_1^{(2k)}(s)\sim [\lambda_e(s)]^{2^k}\, , \qquad b_1^{(2k+1)}(s)\sim [\lambda_o(s)]^{2^k}\, ,  \label{eq:lambda}
\end{equation}
for $k\gg 1$. Dependance of $\lambda_e$ and $\lambda_o$ on values of the bending parameter $s$, obtained by numerical iteration of $\frac{\ln b_1^{(r)}(s)}{2^{[r/2]}}$, is depicted in figure~\ref{fig:omegaLambda}.

Now, using the rescaled variable $b_1^{(r)}$,  the partition function  (\ref{eq:statsum}) may be written as
\begin{equation}
Z_c^{(r+1)}=\left(b_1^{(r)}E_1^{(r)}\right)^2\, ,
\end{equation}
so that, introducing new variables
\begin{equation}
y_r=\frac{\ln Z_c^{(r)}}{N_r}\, , \qquad q_r=\frac{\ln E_1^{(r)}}{N_r}\, , \label{eq:defyq}
\end{equation}
where $N_r=2^{r+1}$, one obtains
\begin{equation}
y_{r+1}=q_r+\frac{\ln b_1^{(r)}}{2^{r+1}}\, ,  \label{eq:yq}
\end{equation}
\begin{equation}
q_{r+1}=q_r+\frac{\ln d_2^{(r)}}{2^{r+1}}\, ,  \label{eq:q}
\end{equation}
which follows from the recursion relation for $E_1^{(r)}$ (given in (\ref{eq:recur})) and definition (\ref{eq:skalirani}). Numerically iterating recursion equation for $q_r$, for various values of $s$, one obtains that finite limiting value $\lim_{r\to\infty} q_r$ exists and it depends on $s$. Then, from (\ref{eq:yq}) and (\ref{eq:lambda}) it follows that
\begin{equation}
\lim_{r\to\infty} y_r=\lim_{r\to\infty} q_r =\ln\omega(s)\, , \label{eq:omega}
\end{equation}
meaning that the leading factor in asymptotical behavior of $Z_c^{(r)}$ is $\omega^{N_r}$. Values of $\omega(s)$ are depicted in figure~\ref{fig:omegaLambda}.
To find the next term in the asymptotical formula for $\ln Z_c^{(r)}$, we observe that, using (\ref{eq:yq}), (\ref{eq:omega}) and (\ref{eq:q}), one  obtains
\begin{equation}
y_{r+1}=\ln\omega+\frac{\ln b_1^{(r)}}{2^{r+1}}-\sum_{i=r}^{\infty}(q_{i+1}-q_i)=\ln\omega+\frac{\ln b_1^{(r)}}{2^{r+1}}-\sum_{i=r}^{\infty}\frac{\ln d_2^{(i)}}{2^{i+1}}\, .\end{equation}
Taking into account that $|\ln d_2^{(i)}|$ is less than some finite constant (which was numerically obtained), as well as (\ref{eq:lambda}), one can conclude that for $r\gg 1$ the following approximate relation follows
\begin{equation}
y_{r+1}\approx \ln\omega +\frac{\ln b_1^{(r)}}{2^{r+1}}\, ,
\end{equation}
which implies
\begin{equation}
Z_c^{(r)}(s)\sim [\omega(s)]^{N_r}\times\cases{[\mu_e(s)]^{\sqrt{N_r}},& for $r$ even\\
 [\mu_o(s)]^{\sqrt{N_r}},& for $r$ odd\\}  \qquad ,
                                        \end{equation}
where $\mu_e(s)=[\lambda_o(s)]^{1/\sqrt 2}$ and $\mu_o(s)=\lambda_e(s)$.

\subsection{Thermodynamics of semi-flexible Hamiltonian cycles on $p=2$ MR lattice}
\label{treca}

\begin{figure}%[b]
\begin{center}
\includegraphics[scale=0.4]{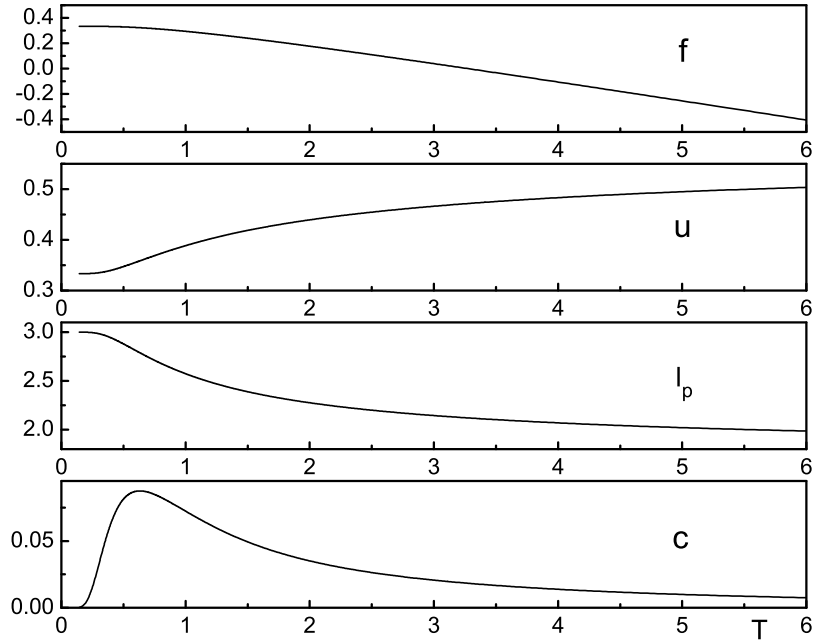}
\end{center}
\caption{Free energy  $f$ (\ref{eq:slobodnaEnergija}), internal energy $u$ (\ref{eq:unutrasnja}), persistence length $l_p$ (\ref{plength}), and heat capacity $c$ (\ref{eq:kapacitet}) per monomer, in the thermodynamic limit,  as functions of temperature $T$ ($f$ and $u$ are measured in units of $\varepsilon$, $c$ in units of $k_B$, and $T$ in units of $\varepsilon/k_B$), for $p=2$ MR lattice.}
\label{fig:termodinamika}
\end{figure}

By definition, the free energy per monomer, in the thermodynamic limit, is equal to
\begin{equation}\label{fdef}
   f=-k_BT\lim_{r\rightarrow\infty}\frac{\ln Z_c^{(r)}}{N_r}\,,
\end{equation}
so that, from (\ref{eq:defyq}) and (\ref{eq:omega}), it follows
\begin{equation}
f=-k_BT\ln\omega=\varepsilon\frac{\ln\omega}{\ln s}\, .   \label{eq:slobodnaEnergija}
\end{equation}
Using already found values of $\omega(s)$ (depicted in figure~\ref{fig:omegaLambda}), one can obtain $f(T)$, which is shown in figure~\ref{fig:termodinamika}. One can see that $f$ is a differentiable function of $T$.

Internal energy per monomer, in the thermodynamic limit, is equal to
\begin{equation}\label{udef}
   u=\varepsilon\lim_{r \to \infty}\frac{\langle
N_b^{(r)}\rangle}{N_r}=s\frac{\partial}{\partial s}\left(f\ln s\right)\,,
\end{equation}
where $N_b^{(r)}$ is the number of bends within the HW. Using (\ref{eq:slobodnaEnergija}) and (\ref{eq:omega}), one obtains
\begin{equation}
\frac u{\varepsilon}=s\frac{\partial}{\partial s}(\ln\omega)=s\,\lim_{r\to\infty} q_r'\, ,  \label{eq:unutrasnja}
\end{equation}
where prime denotes derivative of $q_r$ with respect to $s$.  The recursion   relation  for $q_r'$ follows  from relation (\ref{eq:q}) and has the form
\begin{equation}
q_{r+1}'=q_r'+\frac 1{N_r}\frac{\left(d_2^{(r)}\right)'}{d_2^{(r)}}\, ,  \label{eq:racizvodi}
\end{equation}
whereas from (\ref{eq:malerec}) one can directly obtain recursion relations for derivatives of $x^{(r)}$ (defined by (\ref{eq:skalirani})). Iterating all these relations, internal energy $u$ can be calculated for any particular $s$.

Persistence length is  defined as an average number of steps between two consecutive bends
\begin{equation}
l_p= \lim_{r \to \infty}\frac{N_r}{\langle
N_b^{(r)}\rangle}=\frac{\varepsilon}{u}\, ,  \label{plength}
\end{equation}
and can be evaluated directly from $u$.

Using expressions obtained for $u$, one can show that the heat capacity per monomer $c=\frac{\partial u}{\partial T}$ is equal to
 \begin{equation}\label{c}
 c=k_B\ln^2s\left[\frac u\varepsilon +s^2 \frac{\partial^2}{\partial s^2}(\ln\omega)\right]\, .  \label{eq:kapacitet}
\end{equation}
Since $\ln\omega=\lim_{r\to\infty} q_r$, this means that in order to calculate heat capacity, in addition to already calculated $u$, one needs second derivatives of $q_r$, for $r\gg 1$. These derivatives can be obtained recursively using the relation
\begin{equation}
q_{r+1}''=q_r''+\frac 1{N_r}\left[\frac{\left(d_2^{(r)}\right)''}{d_2^{(r)}}-\left(\frac{\left(d_2^{(r)}\right)'}{d_2^{(r)}}\right)^2\right]\, , \label{eq:racizvodi2}
\end{equation}
which follows directly from  (\ref{eq:racizvodi}), together with recursion relations (\ref{eq:malerec}) for $x^{(r)}$ and corresponding recursive relations for their first and second derivatives, which can be obtained straightforwardly. Temperature dependance of all evaluated thermodynamic quantities is depicted in figure~\ref{fig:termodinamika}, whereupon one can perceive that the free energy $f$ and the persistence length of the polymer  monotonically decrease with $T$, whereas the internal energy $u$ is monotonically increasing function of $T$. Finally, the specific heat $c$ is a non-monotonic function of temperature, displaying a maximum for some $T<1$ (in the units of $\varepsilon/k_B$).

%One should note here that values of $q_r'$ and $q_r''$ saturates quickly with $r$.

\section{Generalization to  MR lattices with $p>2$}
\label{cetvrta}

\begin{figure}%[b]
\begin{center}
\includegraphics[scale=0.65]{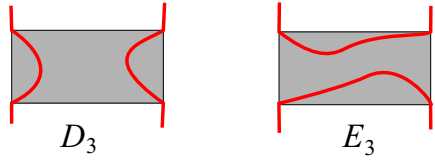}
\end{center}
\caption{Two additional types of semi-flexible HWs on the $r$th order fractal structure, for any $p>2$ MR lattice. Other possible  conformations are of the same type as for  $p=2$ MR lattice, and they are  depicted in figure~\ref{fig:redukovane}.}
 \label{fig:redukovanep}
\end{figure}

It is straightforward to generalize the method for lattices with $p>2$. Due to the  connectivity of the lattices  and symmetry considerations, it follows  that for  any $p>2$ there can be altogether eleven possible types of semi-flexible   conformations. The nine ones, shown in figure~\ref{fig:redukovane},  have already been introduced in the case of  $p=2$.    Two   additional  ones  needed  in the case of  $p>2$ are  shown in figure~\ref{fig:redukovanep}.
For general  $p>2$  restricted partition functions of  these conformations  satisfy the following recursion equations
\begin{eqnarray}
 \fl A_1^{(r+1)}=B_1^{(r)}D_1^{(r)}\left(D_3^{(r)}\right)^{p-2}\, , \qquad A_2^{(r+1)}=B_1^{(r)}D_2^{(r)}\left(D_3^{(r)}\right)^{p-2}\, ,   \nonumber\\
 \fl B_1^{(r+1)}=\left(A_1^{(r)}\right)^{p-2}\left(A_2^{(r)}\right)^2\, , \qquad B_2^{(r+1)}=\left ( A_1^{(r)}\right)^{p-1}A_2^{(r)}\, , \qquad B_3^{(r+1)}=\left( A_1^{(r)}\right)^{p}\, , \nonumber \\
 \fl D_1^{(r+1)}=
 2D_2^{(r)}\left(D_3^{(r)}\right)^{p-2}E_2^{(r)}+
 (p-2)\left(D_2^{(r)}\right)^2\left(D_3^{(r)}\right)^{p-3}E_3^{(r)}\nonumber \\
  \! \! \!\! \! \!\! \! \!\! \! \!+2B_1^{(r)}B_2^{(r)}D_2^{(r)}\left(D_3^{(r)}\right)^{p-3}
  +(p-3)\left(B_1^{(r)}\right)^2\left(D_2^{(r)}\right)^2\left(D_3^{(r)}\right)^{p-4}\, ,  \nonumber\\
\fl D_2^{(r+1)}=D_2^{(r)}\left(D_3^{(r)}\right)^{p-2}E_1^{(r)}+
D_1^{(r)}\left(D_3^{(r)}\right)^{p-2}E_2^{(r)}+
(p-2)D_1^{(r)}D_2^{(r)}\left(D_3^{(r)}\right)^{p-3}E_3^{(r)}\nonumber\\
 \! \! \!\! \! \!\! \! \!\! \! \!+B_1^{(r)}B_3^{(r)}D_2^{(r)}\left(D_3^{(r)}\right)^{p-3}+
B_1^{(r)}B_2^{(r)}D_1^{(r)}\left(D_3^{(r)}\right)^{p-3}\nonumber\\
 \! \! \!\! \! \!\! \! \!\! \! \!+(p-3)\left(B_1^{(r)}\right)^2D_1^{(r)}D_2^{(r)}\left(D_3^{(r)}\right)^{p-4}\, ,  \nonumber\\
 \fl D_3^{(r+1)}=2D_1^{(r)}\left(D_3^{(r)}\right)^{p-2}E_1^{(r)}+
 (p-2)\left(D_1^{(r)}\right)^2\left(D_3^{(r)}\right)^{p-3}E_3^{(r)}\nonumber \\
 \! \! \!\! \! \!\! \! \!\! \! \!   +2B_1^{(r)}B_3^{(r)}D_1^{(r)}\left(D_3^{(r)}\right)^{p-3}
  +(p-3)\left(B_1^{(r)}\right)^2\left(D_1^{(r)}\right)^2\left(D_3^{(r)}\right)^{p-4}\, , \nonumber\\
\fl E_1^{(r+1)}=\left(D_2^{(r)}\right)^2\left(D_3^{(r)}\right)^{p-2}\, , \qquad\qquad  E_2^{(r+1)}=D_1^{(r)}D_2^{(r)}\left(D_3^{(r)}\right)^{p-2}\, , \nonumber \\
 \fl E_3^{(r+1)}=
\left(D_1^{(r)}\right)^2\left(D_3^{(r)}\right)^{p-2}\, ,\label{eq:recurp}
\end{eqnarray}
with the initial values given on the  unit square, which are  for the new variables given by  $D_3^{(1)}=1$ and $E_3^{(1)}=s^4$, while for the other variables  they are the same as for the $p=2$ case.   Partition function of  all closed semi-flexible conformations on the generator of order $(r+1)$, for  an  arbitrary $p>2$ member of MR family, can be written as
\begin{equation}\label{eq:statsump}
  Z_c^{(r+1)}=\left(B_1^{(r)}\right)^2\left(D_3^{(r)}\right)^{p-2}\, .
\end{equation}

As in the case of $p=2$ MR fractal, it is convenient to  rescale the set of variables $X \in \{A_1, A_2, B_1, B_2, B_3, D_1, D_2, D_3, E_2, E_3\}$ by dividing them with the variable $E_1$, so that the new ones
\begin{equation}
x^{(r)}=\frac{X^{(r)}}{E_1^{(r)}}\, , \qquad x\in \{a_1, a_2, b_1, b_2, b_3, d_1, d_2, d_3, e_2, e_3\}\, ,  \label{eq:skaliranip}
\end{equation}
obey  the recurrence equations
\begin{eqnarray}
\fl a_1^{(r+1)}=\frac{b_1^{(r)}d_1^{(r)}}{\left(d_2^{(r)}\right)^2}\, , \qquad a_2^{(r+1)}=\frac{b_1^{(r)}}{d_2^{(r)}}\, ,  \qquad b_1^{(r+1)}=\left(\frac{a_2^{(r)}}{d_2^{(r)}}\right)^2\left(\frac{a_1^{(r)}}{d_3^{(r)}}\right)^{p-2}\, ,\nonumber\\
\fl b_2^{(r+1)}=\frac{a_1^{(r)}a_2^{(r)}}{\left(d_2^{(r)}\right)^2}
\left(\frac{a_1^{(r)}}{d_3^{(r)}}\right)^{p-2}\, , \qquad \qquad b_3^{(r+1)}=\left(\frac{a_1^{(r)}}{d_2^{(r)}}\right)^2\left(\frac{a_1^{(r)}}{d_3^{(r)}}\right)^{p-2}\, , \nonumber\\
\fl d_1^{(r+1)}=2\frac{e_2^{(r)}}{d_2^{(r)}}+(p-2)\frac{e_3^{(r)}}{d_3^{(r)}}+2\frac{b_1^{(r)}b_2^{(r)}}{d_2^{(r)}d_3^{(r)}}+(p-3)\left(\frac{b_1^{(r)}}{d_3^{(r)}}\right)^2\, , \nonumber \\
\fl d_2^{(r+1)}=\frac{1}{d_2^{(r)}}+\frac{d_1^{(r)}e_2^{(r)}}{\left(d_2^{(r)}\right)^2}
+\frac{(p-2)d_1^{(r)}e_3^{(r)}+b_1^{(r)}b_3^{(r)}}{d_2^{(r)}d_3^{(r)}}
+\frac{b_1^{(r)}b_2^{(r)}d_1^{(r)}}{\left(d_2^{(r)}\right)^2d_3^{(r)}}+
(p-3)\frac{(b_1^{(r)})^2d_1^{(r)}}{d_2^{(r)}\left(d_3^{(r)}\right)^2}\, , \nonumber \\
\fl d_3^{(r+1)}=2\frac{d_1^{(r)}}{\left(d_2^{(r)}\right)^2}+
(p-2)\frac{(d_1^{(r)})^2e_3^{(r)}}{\left(d_2^{(r)}\right)^2d_3^{(r)}}+
2\frac{b_1^{(r)}b_3^{(r)}d_1^{(r)}}{\left(d_2^{(r)}\right)^2d_3^{(r)}}+
(p-3)\frac{(b_1^{(r)})^2(d_1^{(r)})^2}{\left(d_2^{(r)}\right)^2\left(d_3^{(r)}\right)^2}\, ,\nonumber \\
\fl e_2^{(r+1)}=\frac{d_1^{(r)}}{d_2^{(r)}}\, , \qquad\qquad \qquad e_3^{(r+1)}=\left(\frac{d_1^{(r)}}{d_2^{(r)}}\right)^2 \, ,\label{eq:malerecp}
\end{eqnarray}
which follow from (\ref{eq:recurp}), whereas the
equation for  $E_1$  becomes
\begin{equation}\label{eq:E1}
E_1^{(r+1)}=\left(d_2^{(r)}\right)^2\left(d_3^{(r)}\right)^{p-2}\left(E_1^{(r)}\right)^{p}\,,
\end{equation}
so that, the partition function (\ref{eq:statsump}), in new variables gets the form
\begin{equation}\label{eq:statsumnew}
Z_c^{(r+1)}=\left(b_1^{(r)}\right)^2\left(d_3^{(r)}\right)^{p-2}\left(E_1^{(r)}\right)^{p}\, .
\end{equation}

\begin{figure}%[b]
\begin{center}
\includegraphics[scale=0.4]{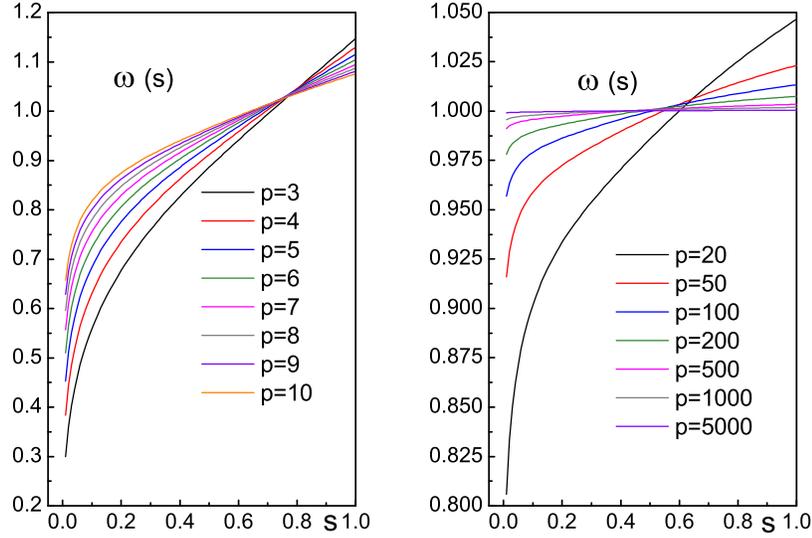}
\end{center}
\caption{ Stiffness dependance of the base $\omega$ in (\ref{eq:statsumfin}), for various members of MR family, labelled by parameter $p$.}
 \label{fig:omegap}
\end{figure}
\begin{figure}%[b]
\begin{center}
\includegraphics[scale=0.4]{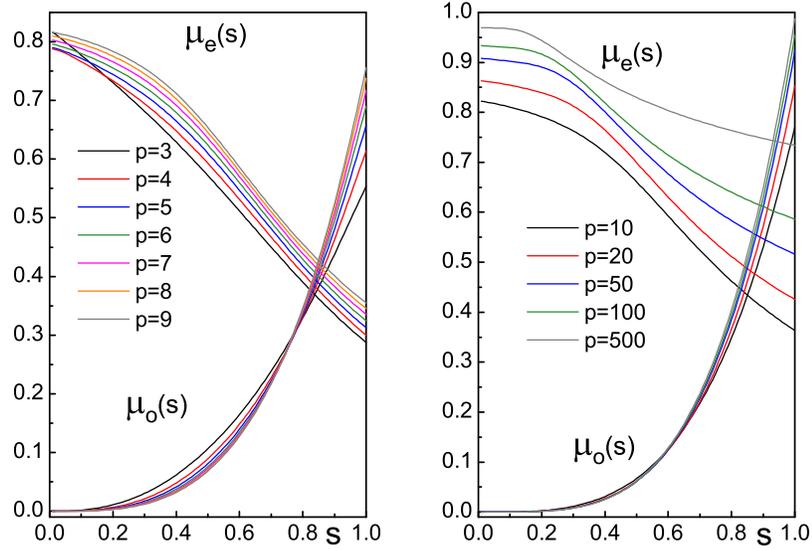}
\end{center}
\caption{Stiffness dependence of the bases $\mu$ in the stretched exponential factor in (\ref{eq:statsumfin}), for various members of MR family, labelled by parameter $p$.}
 \label{fig:mip}
\end{figure}
Iterating the above recursion relations, one can find that  all variables   $a$ and $b$ tend to zero, while variables  $d$ and $e$ tend to some finite constants, which depend on the parity of the generator order.  The trend is  such that  this convergence is faster  on fractals with    higher values of $p$. For arbitrary $p$ we find,  similarly to  equations (\ref{eq:lambda}), that  $b_1$  approaches zero as

\begin{equation}
b_1^{(2k)}(s)\sim [\lambda_{e}(s)]^{p^k}\, , \qquad b_1^{(2k+1)}(s)\sim [\lambda_{o}(s)]^{p^k}\, ,  \label{eq:lambdap}
\end{equation}
where constants $\lambda_e$ and $\lambda_o$ depend on the fractal  parameter $p$.
Following the same procedure as in  subsection \ref{metodsekcija}, for the asymptotic behavior of the partition function,  for general $p$ we again obtain the scaling form
\begin{equation}\label{eq:statsumfin}
Z_c^{(r)}(s)\sim [\omega(s)]^{N_r}\times\cases{[\mu_e(s)]^{\sqrt{N_r}},& for $r$ even\\
 [\mu_o(s)]^{\sqrt{N_r}},& for $r$ odd\\}  \qquad ,
 \end{equation}
where now $\mu_e(s)=[\lambda_o(s)]^{1/\sqrt p}$ and $\mu_o(s)=\lambda_e(s)$. Dependance of $\omega$ on the stiffness parameter $s$,  for various values of $p$, is given in figure~\ref{fig:omegap}, where one can observe that for very large $p$ the quantity $\omega$ approaches the unit value, ceasing to depend on $s$. Also, one may notice  that  $\omega(s=1)$  is smaller  for lattices with higher value of  $p$,  meaning that  the number of fully flexible HWs on equally large lattices is smaller for higher $p$. The reason  for this is that the number of edges, and  therefore connectivity of lattices, decreases with  $p$. Values of $\mu_e$ and $\mu_o$, as functions of $s$, are shown in figure~\ref{fig:mip}, for various values of $p$, where one can see that $\mu_e$ decreases, while $\mu_o$ increases with $s$, for each member of  MR family.

\begin{figure}%[b]
\begin{center}
\includegraphics[scale=0.35]{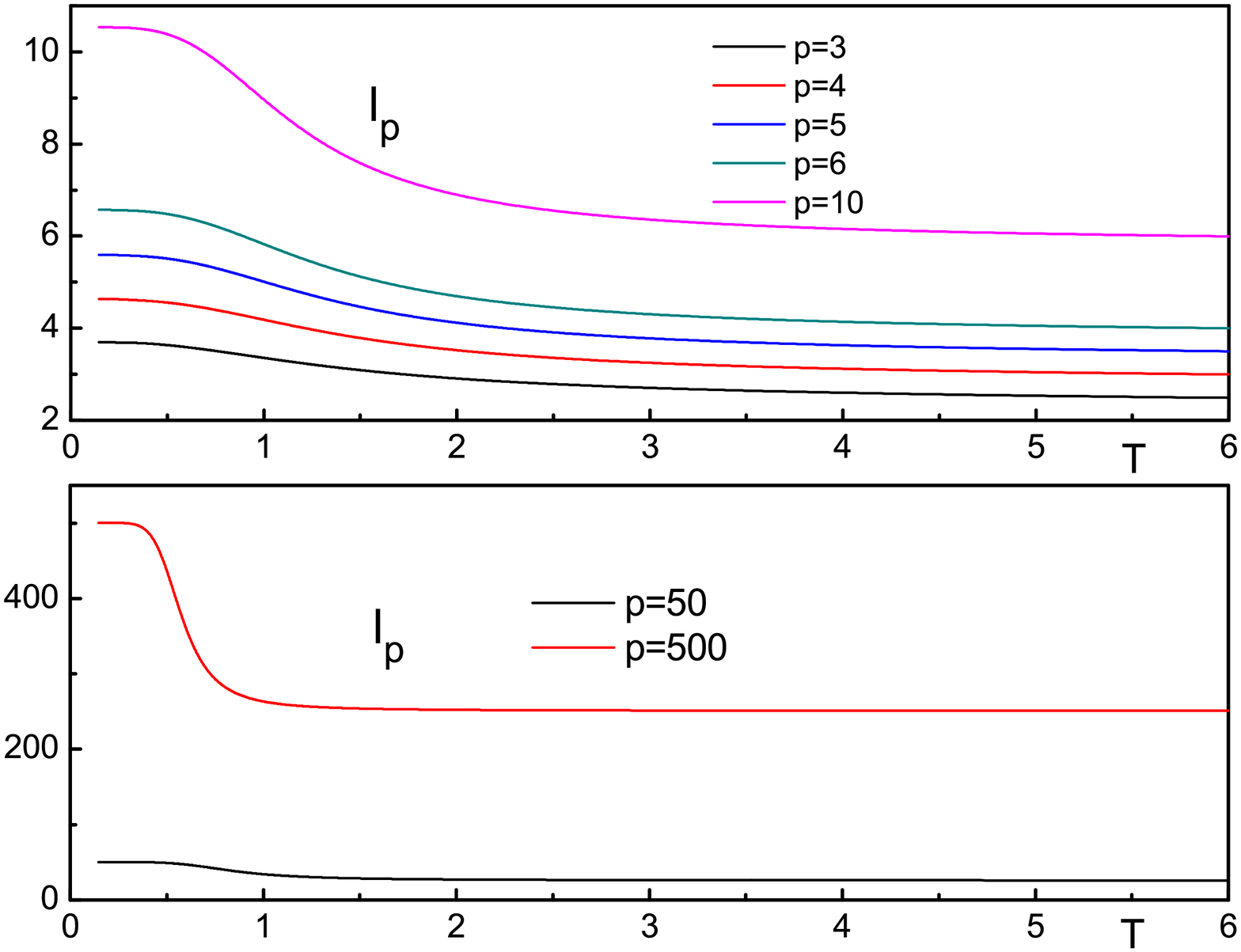}
\end{center}
\caption{Persistence length as a function of temperature, for various values of parameter $p$ that enumerates members of MR family.}
 \label{fig:lp}
\end{figure}
\begin{figure}%[b]
\begin{center}
\includegraphics[scale=0.35]{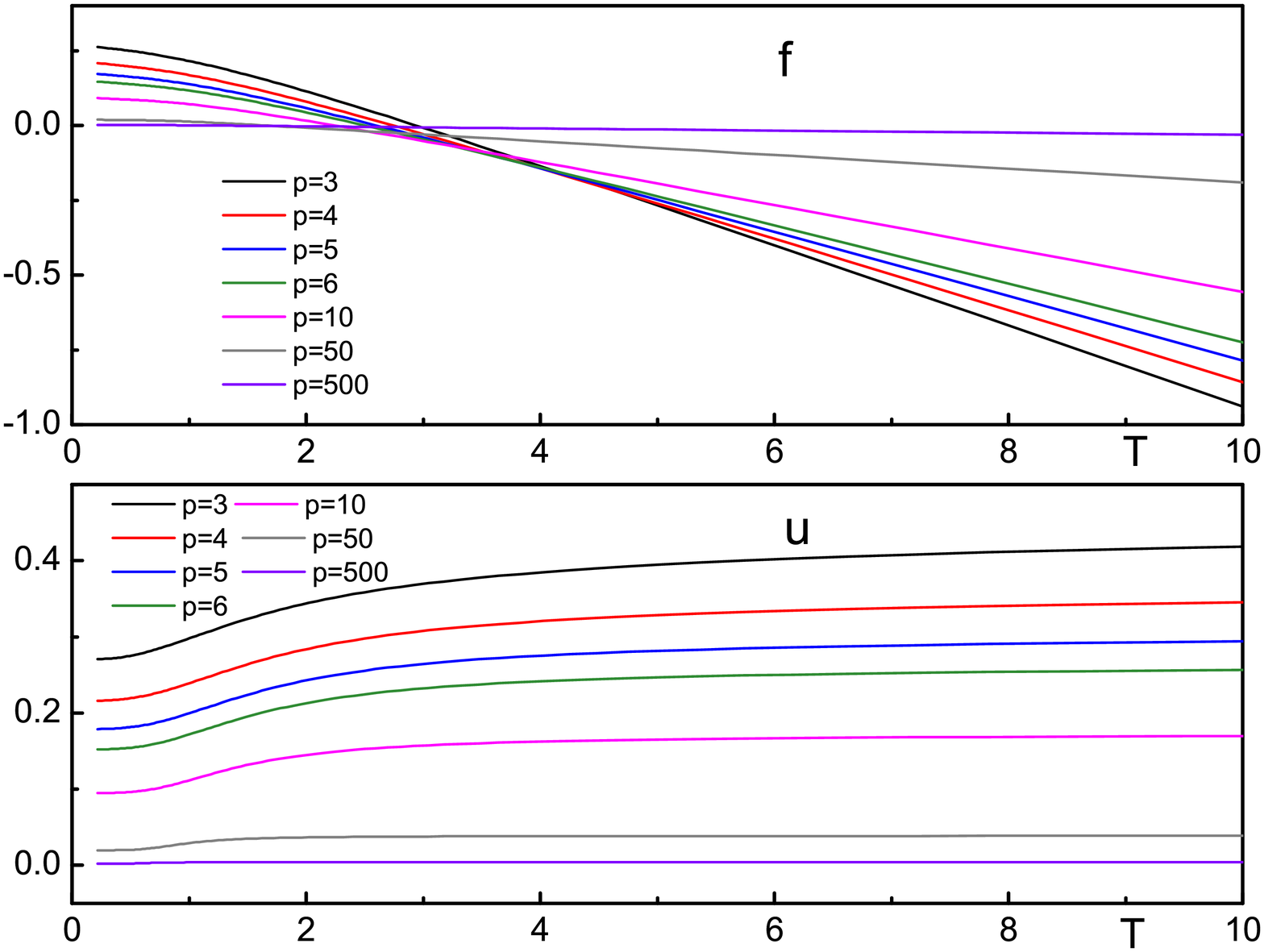}
\end{center}
\caption{Free energy $f$ and internal energy $u$ as functions of temperature, for various values of parameter $p$ that enumerates members of MR family. }
 \label{fig:fp}
\end{figure}
\begin{figure}%[b]
\begin{center}
\includegraphics[scale=0.35]{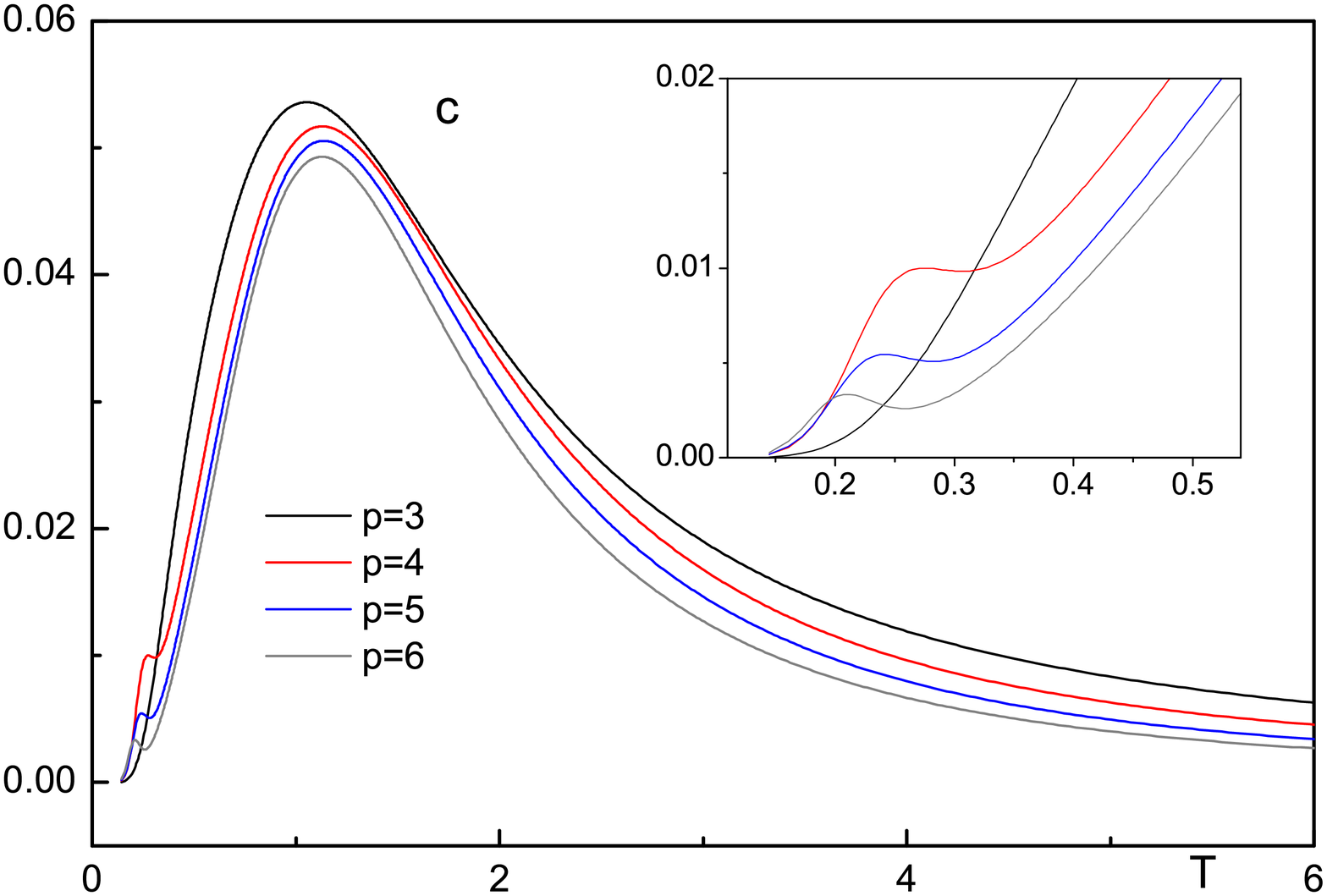}
\end{center}
\caption{Specific heat $c$ as a function of temperature $T$, for  MR fractals labelled by $p=3, 4, 5$, and $6$.   Inset graph  highlights additional small peaks that have appeared for $p=4, 5$ and 6 fractals. }
 \label{fig:cp}
\end{figure}
\begin{figure}%[b]
\begin{center}
\includegraphics[scale=0.35]{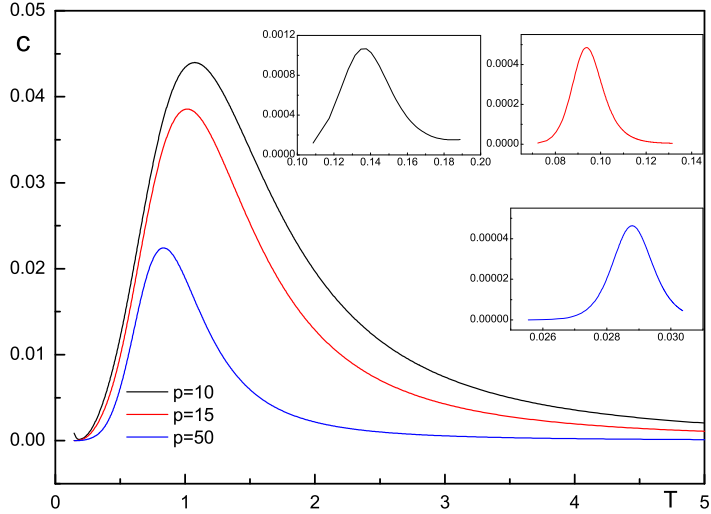}
\end{center}
\caption{Specific heat as function of temperature, for  $p=10$, $p=15$ and $p=50$ fractals.   Inset graphs show small peaks for these fractals. For higher $p$,  peaks are smaller and pulled toward lower temperatures.}
 \label{fig:cpp}
\end{figure}

For $p>2$ thermodynamic functions may be obtained using  the recurrence equations (\ref{eq:malerecp}), and expressions
\begin{equation}\label{eq:qp}
    q_{r+1}=q_r+\frac{1}{4p^{\,r}}\left(2\ln d_2^{(r)}+(p-2)\ln d_3^{(r)}\right)\, ,
\end{equation}
\begin{equation}\label{eq:qpp}
  q_{r+1}'=q_r'+\frac{1}{4p^{\,r}}\left(2\frac{\left(d_2^{(r)}\right)'}{d_2^{(r)}}
  +(p-2) \frac{\left(d_3^{(r)}\right)'}{d_3^{(r)}}\right)\, ,
\end{equation}
\begin{eqnarray}
% \nonumber to remove numbering (before each equation)
  q_{r+1}'' &=& q_r''+\frac{1}{4p^{\,r}}\left[2\frac{\left(d_2^{(r)}\right)''}
  {d_2^{(r)}}-2\left(\frac{\left(d_2^{(r)}\right)'}{d_2^{(r)}}\right)^2\right] \nonumber \\
   &+&\frac{1}{4p^{\,r}}\left[(p-2)\frac{\left(d_3^{(r)}\right)''}{d_3^{(r)}}
   -(p-2)\left(\frac{\left(d_3^{(r)}\right)'}{d_3^{(r)}}\right)^2\right]\,,
\end{eqnarray}
that correspond to the equations (\ref{eq:q}), (\ref{eq:racizvodi}) and (\ref{eq:racizvodi2}) (obtained for $p=2$ case), respectively.

The obtained numerical results for the persistence length $l_p$ as function of temperature $T$, for different MR fractals,  are  depicted in figure~\ref{fig:lp}, where one can see that $l_p$ decreases with temperature, implying that number of polymer bends increases with $T$.
Dependance of free  and internal energy on $T$   is  presented in figure~\ref{fig:fp}, for various members of MR family. One perceives that $f$ monotonically decreases, while $u$ monotonically increases with $T$, for each $p$. Also, in the limit of very large $p$, one can conclude that both $f$ and $u$ go to zero.
The obtained increment of internal energy with temperature  is in accordance with the fact  that at lower temperatures energetic effects dominate, so that low  energy levels with conformations consisting of smaller number of bends are more populated. At higher temperatures, all energy levels become populated and internal energy saturates (i.e. becomes constant). This saturation is faster  on fractals with larger values of $p$, for which the internal energy is generally smaller. The reason for this lies in the connectivity of the  vertices. For lattices oriented as in figure~\ref{fig:mrlattice} there are more vertical  than horizontal edges, and for  lattices with larger value of $p$ this anisotropy becomes larger. The walks follow preferred direction and make smaller number of turns which reduces energy and increases persistence length.  Described   behavior of internal energy implies that  specific heat should have a peak in the low temperature region, which we have numerically confirmed and displayed in  figures~\ref{fig:cp} and \ref{fig:cpp}, where specific heat as a function of $T$ is shown.
In these figures one can notice that besides one pronounced peak in specific heat landscape, there is another small peak at low temperatures, for fractals with $p\geq4$.
This effect in specific heat behaviour is known as Schottky  anomaly (see, for instance \cite{schottky}) and appears in systems with a finite number of energy levels.

We finish our discussion  inferring  that within the studied compact polymer phase there is no finite order phase transition, due to the fact that entropy and specific heat are continuous,
smooth functions of  temperature.  Since the  persistence length $l_p$ is finite at any $T$, the polymer system is always in liquid-like (disordered) phase, and the transition to  the crystal  (ordered) phase is not possible. The existence of only disordered compact phase has also  been observed in the case of semi-flexible HW on 3- and 4-simplex lattices \cite{Phaysica2011}. The absence of crystal phase  on the studied family of lattices stems from their asymmetry  in horizontal and vertical direction. For each MR fractal there are more vertical than horizontal bonds. This discrepancy is more pronounced for larger $p$ lattices, implying  smaller number of bends in compact conformations since they are forced by the lattice in the vertical direction. Nevertheless, on  such lattices  conformations still have  a  large number of horizontal steps  that prevent an ordered state that can exist on square lattice \cite{Jacobsen}.

\section{Ground states and frustration}
\label{peta}

In order to achieve a minimal energy state at $T=0$, in this section only conformations with a minimal number of bends will be considered.  First we analyse  the case of $p=2$ lattice. Since the  conformation $D_2$    makes the smallest number of bends on the unit square, one expects that the ground state, in this case, would be comprised of  HW conformations with the  maximal possible number  of  $D_2$ type  on each unit square. Contribution of the ground state to the  whole partition function is of the  form  $Z_0=N_0s^{N_{b0}}$, with  $N_0$ being  the number of ground state HWs   and $N_{b0}$  being the number of bends in each of these conformations.  This term in partition function can be obtained from  relation  (\ref{eq:statsum}) and  recurrence equations (\ref{eq:recur}), keeping only the terms with conformations of type $D_2$. Then, some of the variables drop, and the system (\ref{eq:recur}) reduces to
\begin{eqnarray}\label{eq:groundp2}
% \nonumber to remove numbering (before each equation)
 A_2^{(r+1)}&=& B_1^{(r)}D_2^{(r)}\, ,\qquad  B_1^{(r+1)}=\left(A_2^{(r)}\right)^2\, , \nonumber\\
 D_2^{(r+1)}&=&D_2^{(r)}E_1^{(r)}\, , \qquad E_1^{(r+1)}=\left(D_2^{(r)}\right)^2\, .
\end{eqnarray}
Solving this system exactly, from (\ref{eq:statsum}) we obtained $Z_0$ on  the $r$th order fractal structure
\begin{equation}\label{ew:stat0}
Z_{0r}=s^{(\sqrt{2})^{r+1}[1+(-1)^{r+1}]+\frac{2}{3}\,2^r+\frac{4}{3}(-1)^r}\,.
\end{equation}
In this case  the  ground state is non-degenerate, with the only one conformation leading to the zero entropy.    From the   number of bends in this conformation,  given by $N_{b0r}=(\sqrt{2})^{r+1}[1+(-1)^{r+1}]+\frac{2}{3}\,2^r+\frac{4}{3}(-1)^r$, we could calculate ground state energy per site in the thermodynamic limit, as $u_0=\varepsilon\lim_{r \to \infty}\frac{N_{b0r}}{2^{r+1}}$. The obtained value is $\frac{u_0}{\varepsilon} =\frac{1}{3}$, which is verified numerically and can be seen in figure~\ref{fig:termodinamika}.
\begin{figure}%[b]
\begin{center}
\includegraphics[scale=0.25]{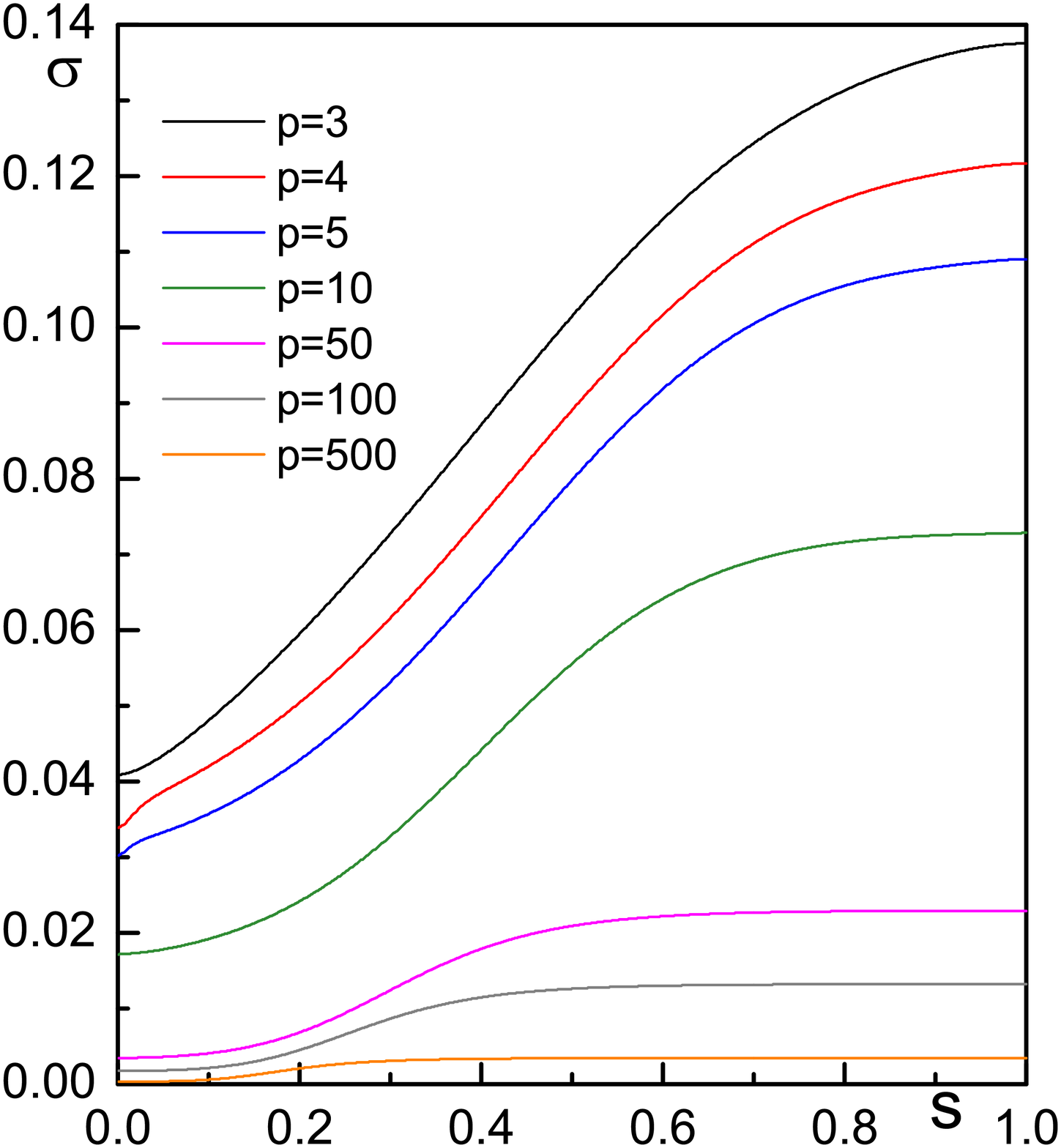}
\end{center}
\caption{Entropy per monomer $\sigma$, in thermodynamic limit, as a function of the stiffness parameter $s$, for various members of MR fractal family (labelled by $p$).}
 \label{fig:en3}
\end{figure}

For  $p>2$  equations are more complicated, and we have not been able to extract exact expressions for the number of ground state conformations, but numerically  we have calculated  the entropies per site, in the thermodynamic limit, in the whole range of stiffness parameter $s$  (see figure~\ref{fig:en3}). One can observe that for $p>2$ fractals,  ground state entropies per monomer do not vanish, meaning that there are exponentially large number of ground state conformations, which is a characteristic of  geometrically frustrated systems. Limiting values of  entropies for  various MR fractals are given in
table~\ref{tab:en3}.
\begin{table}
\caption{Entropies  per monomer $\sigma^*$  at temperature $T=0$, for various $p$ fractals of MR family. We see that $\sigma^*$ (and consequently the number of ground state conformations)  decreases  with $p$.}
\centering
\label{tab:en3}
\vspace{2mm}
\begin{tabular}{c|ccccccc}
\hline
% after \\: \hline or \cline{col1-col2} \cline{col3-col4} ...
${ p}$ &3 & 4 & 5 & 10 & 50 & 100 &500 \\ \hline
${\sigma^*}$ & 0.040902 & 0.033925 & 0.030247 & 0.016914 & 0.0034679 & 0.0017333 & 0.00034658 \\  \hline
 \end{tabular}
\end{table}

\section{Summary and conclusion}
\label{sesta}

We have studied a model of compact semi-flexible polymer rings modelled by closed Hamiltonian walks on the family of MR fractal lattices, whose members are labelled by an integer $p\geq2$. All lattices from the family have the same fractal dimension ($d_f=2$) and the coordination number  (three), but their vertices are connected  differently. Lattices  can be obtained from the square lattice by deleting some bonds from  it, which induces anisotropy between  horizontal  and vertical direction. By applying an exact method of recurrence relations, we have established the scaling form of the corresponding partition function  (given by equation (\ref{eq:statsumfin})) on  the whole family of fractals. There is a leading exponential factor with a base  $\omega$, which depends on the lattice parameter $p$, as well as on the stiffness parameter $s$. For each $p$ studied, we have found numerically  that $\omega$ is increasing function of $s$, and that it changes more slowly on fractals with higher $p$. Correction to the leading exponential factor is stretched exponential factor of the same form for each fractal of the considered family, in the whole range of $s$ values.

From the obtained partition function we have  evaluated the set of  thermodynamic quantities (free and internal energy, specific heat and entropy) as well as the polymer persistence length, as functions of the stiffness parameter $s$ (or temperature $T$). We have found that all these  quantities are differentiable functions of  $s$. For each member of MR family, we have found that all these quantities are monotonic functions of $T$, except for the specific heat, which has a maximum at low temperatures. Since the entropy and specific heat are continuous, smooth functions of  temperature, there is no finite order phase transition, and the studied polymer system  can  exist only in disordered phase.

Eventually, we  have analysed the ground state of the studied model.  For $p=2$ fractal we have found that the ground state  is non-degenerate, and that the only ground state conformation has the persistence length $l_p=3$. So, on average,  there is one bend after every three steps, and there  are no long straight segments in this conformation. The  number  of left/right and up/down turns are comparable and this conformation is disordered. On the other hand, for  fractals  with $p>2$, the ground state is degenerate, with  exponentially large number of conformations, producing the residual entropy.  The  number of ground state conformations is maximal for $p=3$ and decreases with $p$. Persistence length  is the smallest for $p=3$ ground state, and becomes larger, for larger $p$. However, all these ground state conformations have many bends and do not represent ordered ground states. In fact, we have geometrically frustrated systems, where  geometry of the lattices is in conflict with the condition for minimal energy (i.e. minimal number of bends) and  the requirement that all vertices  are occupied only once. Geometric frustration  suppresses ordered ground states and possibility of ordered phase at any $T$. The studied model describes disordered, liquid-like  compact phase of semi-flexible polymers. Although MR lattices have some resemblance to the square lattice (on  which the ordered phase can exist), an anisotropy  of vertical  and horizontal directions (small for $p=2$, and greater for $p>2$),  causes that ordered  phase can not exist on these lattices.

In conclusion, we may say that the family of modified rectangular lattices proved to be very suitable for an exact recurrence relation study of  conformational properties of  semi-flexible compact polymers in two dimensional nonhomogeneous medium. In our study compact configurations have been described by closed Hamiltonian walks, but this approach can be extended to more complex case of open Hamiltonian walks.  Also, it could  be of  practical significance  to expand the study of examined model into a more realistic case, when polymers are situated in three-dimensional fractal space.

\ack{This paper has been done as a part of the
work within the project No. 171015, funded by  the Ministry of Education, Science and Technological Development of the Republic of Serbia.
}
%%%%%%%%%%%%%%%%%%%%%%%%%%%%%%%%%%%%%%%%%

%\appendix

\end{document}